\documentclass[%
 reprint,
 amsmath,amssymb,
 aps,
prl,
]{revtex4-1}

\usepackage{graphicx}
\usepackage{dcolumn}
\usepackage{bm}
\usepackage[dvipsnames]{xcolor}

\begin{document}

\preprint{APS/123-QED}

\title{Edge dislocations in multi-component solid solution alloys: Beyond traditional elastic depinning}
\author{A. Esfandiapour$^1$, S. Papanikolaou$^1$,  M. Alava$^{1,2}$}
 \affiliation{$^1$NOMATEN Centre of Excellence, National Centre for Nuclear Research, ul.~A. Soltana 7, 05-400 Swierk/Otwock, Poland\\
 $^2$Aalto University, Department of Applied Physics, PO Box 11000, 00076 Aalto, Finland\\
 }
\date{\today}

\begin{abstract}
High-entropy alloys (HEA) form solid solutions with large chemical disorder and excellent mechanical properties. We investigate the origin of HEA strengthening in face-centered cubic (FCC) single-phase HEAs through molecular dynamics simulations of dislocations, in particular, the equiatomic $\rm CrCoNi$, $\rm CrMnCoNi$, $\rm CrFeCoNi$, $\rm CrMnFeCoNi$, $\rm FeNi$, and also, $\rm Fe_{0.4}Mn_{0.27}Ni_{0.26}Co_{0.05}Cr_{0.02}$, $\rm Fe_{0.7}Ni_{0.11}Cr_{0.19}$. The dislocation correlation length $\xi$, roughness amplitude $R_{a}$, and stacking fault widths $W_{SF}$ are tracked as a function of stress. All alloys are characterized by a well defined depinning stress ($\sigma_c$) and we find a novel regime where exceptional strength is observed, and a fortuitous combination takes place, of small stacking fault widths and large dislocation roughness $R_{a}$. Thus the depinning of two partials seems analogous to unconventional domain wall depinning in disordered magnetic thin films. This novel regime is identified in specific compositions commonly associated with exceptional mechanical properties ($\rm CrCoNi$, $\rm CrMnCoNi$, $\rm CrFeCoNi$, and $\rm CrMnFeCoNi$).  Yield stress from analytical solute-strengthening models underestimates largely the results in these cases. A possible strategy for increasing strength in multi-component single-phase alloys is the combined design of stacking fault width and element-based chemical disorder.
\end{abstract}

\maketitle

\begin{figure}[tbh]
	\centering
	\includegraphics[width=\linewidth]{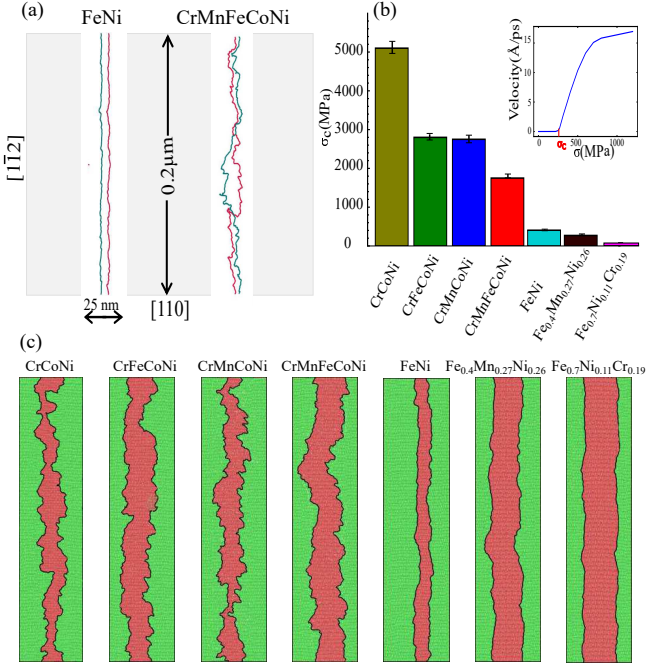}
	\caption{(a) Two partial dislocation lines dissociated from an edge
dislocation for equimolar FeNi and CrMnFeCoNi alloys during flow. Dislocation line direction [1-12] and burger vector direction [110] as well as the size of the box in these directions are presented. (b)  $\sigma_{c}$ for all the alloys in this study. The inset represents the dislocation velocity as a function of $\sigma$, and beyond $\sigma\geq\sigma_{c}$ the dislocation keeps moving. (c) Stacking fault area(red) between two partial dislocations at depinning stress for seven SSAs }
	\label{fig:roughness}
\end{figure}
Solid solution strengthening is one of the key strategies to increase the yield stress of crystalline alloys by introducing solutes that pin dislocations through disturbances in the perfect lattice. In an extreme limit of this process, high entropy alloys (HEAs) are composed of four or more nearly equimolar alloying elements, and they display single-phase behavior with outstanding mechanical properties~\cite{varvenne2016theory,noehring2019, gludovatz2014fracture,wu2014temperature}. Crystal plasticity in HEAs is as common as in any crystal~\cite{george2020high}, driven by dislocation dynamics, but with two key additional variables: chemical-induced-disorder lattice misfit and stacking fault width fluctuations. Common analytical models for solid solution strengthening in HEAs have been focused on quantifying the misfit contributions. For traditional alloys, strengthening was modeled by Fleischer~\cite{fleischer1963} and Labusch~\cite{labusch1970}. They utilized the interaction between solute atoms and the pressure field of a dislocation as the core to their models. In these models, interactions related to the atomic size and shear modulus misfits play an important role in that, the larger the difference between  solute and solvent atom sizes, the stronger the pinning of dislocations.
Beyond traditional solid solution alloys (SSA), in HEAs, a mean-field atomistic potential by Varvenne \textit{et al.}\cite{varvenne2016theory,varvenne2016} was used to identify an effective medium alloy as a reference for HEA with the same average mechanical properties. In this way, Varvenne \textit{et al.} calculated the interaction energy between solutes and dislocations and provided a scaling relationship between strengthening and misfit parameters. Nevertheless, the complexity of stacking fault fluctuations has been left unexplored. In order to investigate the possible effects of stacking fault fluctuations, we study the mechanical properties of edge dislocations and their mobility under externally applied stress for seven FCC SSAs. We find that for the top four stronger alloys, mechanical strength is controlled by a fortuitous combination of small stacking fault widths and chemical-disorder-induced large dislocation roughness, causing dislocation partials' overhangs, possibly analogous to unconventional depinning behaviors in disordered magnetic thin films.

Traditional elastic depinning theory~\cite{Fisher:1998fk, papanikolaou2017avalanches} has been long thought to be the core basis of the description of the behavior of single dislocations  in general disordered environments~\cite{papanikolaou2017avalanches,zapperi2001depinning,zhai2019properties}, and more specifically in solid solutions~\cite{peterffy2020,patinet2011} where chemical disorder proliferates. The key prediction of such elastic depinning theories is the onset of a characteristic length scale, the dislocation correlation length $\xi$, below which the dislocation line displays fractal characteristics with non-trivial roughness. The length $\xi$ is further predicted to scale with the applied stress in a power law manner, maximize at yielding (the depinning transition point) and depend mainly on the dislocation line tension and the disorder fields' fluctuation features. Furthermore, in such theories, the yield point is controlled by the disorder fields' maximum strength, analogously to typical solid solution strengthening theories~\cite{varvenne2016theory,noehring2019, gludovatz2014fracture,wu2014temperature}. Nevertheless, traditional elastic depinning theory does not address the added complexity of the dislocation stacking fault, namely the fact that a gliding dislocation is composed of two dislocation partials' lines that glide together and are separated by a high-energy stacking fault, of average width $W_{SF}$. In metallurgy of pure single-component metals, the width $W_{SF}$ is inversely correlated to the material's yield strength~\cite{asaro2006mechanics}, but its role in strengthening of multicomponent metals, when strong disorder is also present, has been unexplored. 

We investigate the interplay of disorder and stacking faults, by investigating a multitude of equiatomic solid solutions through the use of molecular dynamics simulations~\cite{plimpton2007lammps}. The choices of the studied materials are motivated by prior studies that provided benchmarks for traditional depinning behavior ($\rm Fe_{0.7}Ni_{0.11}Cr_{0.19}$~\cite{peterffy2020}, $\rm FeNi$~\cite{patinet2011}), and also by motivating experimental findings on equiatomic multicomponent alloys~($\rm CrCoNi$, $\rm CrMnCoNi$, $\rm CrFeCoNi$, $\rm CrMnFeCoNi$, $\rm Fe_{0.4}Mn_{0.27}Ni_{0.26}Co_{0.05}Cr_{0.02}$) that point towards exceptional strength~\cite{shang2021mechanical,zhang2020short,li2019strengthening,yao2014novel,li2017strong}. 
Equiatomic fcc HEAs with low stacking fault energy showed an excellent balance between strength and ductility, particularly at cryogenic temperatures~\cite{ gludovatz2014fracture,wu2014temperature}. In an experimental study~\cite{wu2014temperature}, it was shown that the yield strength of the alloys has the following order at 77 K: 
 $\rm CrCoNi>CrMnCoNi>CrFeCoNi> CrMnFeCoNi$
  which indicates that the alloys with the most elements are not necessarily the strongest. Using molecular dynamics (MD) simulations, this Letter focuses on the mobility and geometry of edge dislocations in several random HEAs, employing LAMMPS ~\cite{plimpton1995}  and modified embedded atom method (MEAM) interatomic potential~\cite{choi2018}.
Our focus is the depinning behavior of a model configuration of single edge dislocations under shear stress that drive ideal dislocation glide at the low temperature of 5K.

\begin{figure}[tbh]
    \centering
	\includegraphics[width=\linewidth]{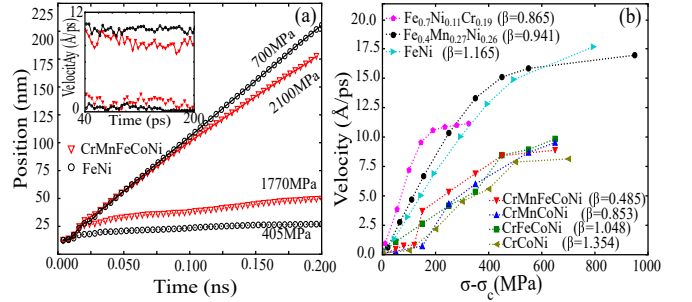}
	\caption{(a) Position and velocity of an edge dislocation as a function of time for equimolar FeNi and CrMnFeCoNi for two different applied stress values. (b)  The velocity of edge dislocation as a function of applied shear stress subtracted by depinning stress ($\sigma_{c}$) for several SSAs. $\beta$ was calculated by fitting these data with a power-law form~\cite{papanikolaou2017avalanches}. }
	\label{fig:velocity}
\end{figure}

\begin{figure}[tbh]
    \centering
    \includegraphics[width=\linewidth]{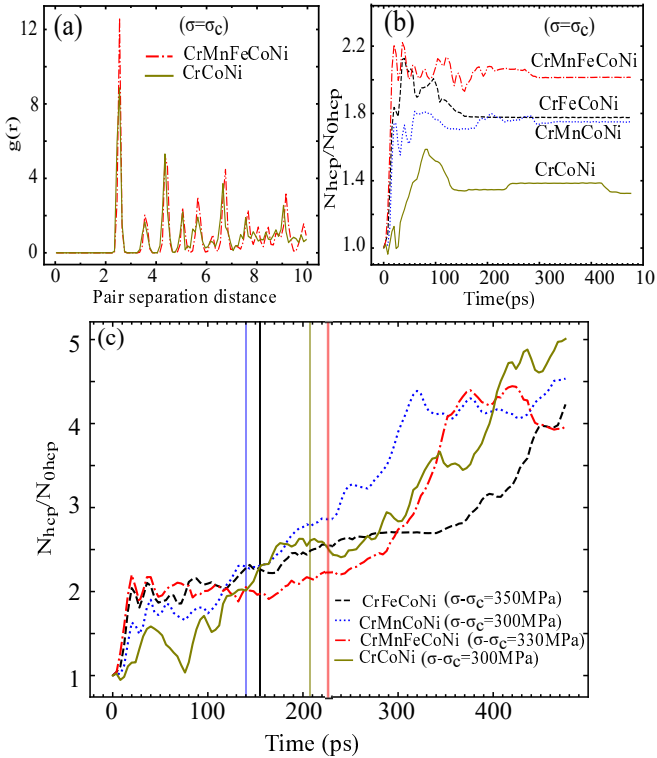}
	\caption{(a) The radial distribution function ($g(r)$) for CrCoNi and CrMnFeCoNi alloys when $\sigma=\sigma_{c}$. The ratio of $N_{hcp}/N_{0hcp}$ as function of time, where $N_{0hcp}$ is the number of hcp atoms at t=0 between tow partial dislocation lines for the four stronger alloys when (b) $\sigma=\sigma_{c}$ and (c) $\sigma>\sigma_c+300 MPa$. The straight lines separate two different regimes for each alloy}.
	\label{fig:hcp}
\end{figure}

 MD simulations can successfully describe the complicated interaction between stacking faults and chemical disorder during loading, at the atomic scale~\cite{peterffy2020,patinet2011}. There are several MD based studies that explained the core structure of dislocations as well as the interaction of dislocations with solutes in fcc traditional alloys~\cite{patinet2011,peterffy2020,zhao2017}. Consistently with these prior studies, our simulations are characterized by a simulation cell with fcc crystal and random distribution of constituent elements, created along $X=[110](l_x=252{\AA})$, $Y=[\bar{1}11](l_y=122{\AA})$, 
and $Z=[1\bar{1}2](l_z=2002{\AA})$ containing 5,432,700 atoms (see Fig.~\ref{fig:roughness}(a)). A periodic array of dislocations (PAD) model~\cite{osetsky2003} was used to insert perfect $\frac{1}{2}\langle110\rangle$ edge dislocation between the two central $[111]$ planes in the cell. Periodic boundary conditions (PBC) are applied in both X and Z direction, while the fixed boundary condition is used in the Y direction. Volume along Y direction is divided into three regions, where the central region contains usual MD mobile atoms and is sandwiched between the fixed upper and lower regions of several atomic layers. First, atomic relaxation is performed using the NPT ensemble to ensure that the stresses in X and Z directions are minimized. Then stress-controlled loading is considered where the force $F_x=\sigma AN_\pm e_{xz}$ is applied to the upper (+) and lower regions (-) with the area of A and $N\pm$ atoms. The simulations are performed in the NVE ensemble with the temperature-controlled by a Berendsen thermostat at 5 K~\cite{berendsen1984}. A time step of 4 fs is used. All seven elemental compositions are simulated up to 300-600 MPa above depinning stress (10-20 different stress values). Each alloy composition is realized three different times. Dislocation and crystal structures are analyzed using dislocation extraction algorithm (DXA)~\cite{stukowski2012dxa} and common neighbor analysis that are implemented in OVITO software~\cite{stukowski2009ovito}. Beyond the apparent computational intensity of this work, it is worth mentioning that the results of this work are based on the analysis of more than 0.5PB of atomic configurational data that is locally stored, given that the dynamics of dislocations is tracked for 10-20 loading stresses at many time steps ($~120$) for every alloy.

\begin{figure}[bth]
	\centering
	\includegraphics[width=\linewidth]{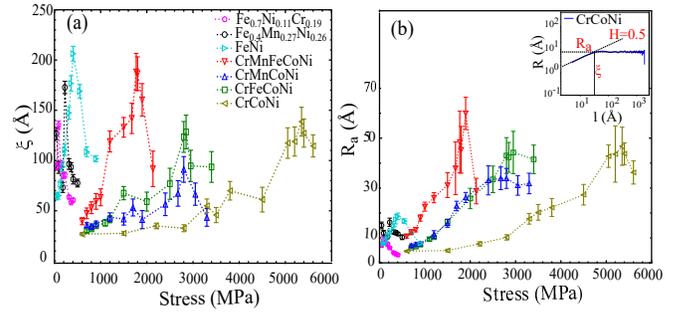}
	\caption{(a) Mean value of correlation length ($\xi$) and (b) Saturation roughness ($R_{a}$) of two partial dislocations which dissociated from an edge dislocation in different HEAs as a function of applied shear stress where the value $H=0.5$ is fixed.}
	\label{fig:roughness2}
\end{figure}

The characteristics of the geometry and mobility of edge dislocations in different HEAs are shown in  Fig.~\ref{fig:roughness} and Fig.~\ref{fig:velocity}.  Fig.~\ref{fig:roughness}(a) has a dislocation line direction [1-12] and burger vector direction along [110].  The figure shows two Shockley partial dislocations in their glide plane, which are dissociated from an edge dislocation for equimolar FeNi and CrMnFeCoNi alloys when $\sigma$ has a large value of 2000 MPa. The overhang of partial dislocations in FeNiCoCrMn can be seen in this figure. Fig.~\ref{fig:roughness}(b) shows the depinning stress ($\sigma_{c}$) identified for seven HEAs. $\sigma_{c}$ is the stress at which the dislocation keeps moving (see the Inset of Fig.~\ref{fig:roughness}(b)). Based on $\sigma_{c}$, we identify two classes of alloys, the four stronger (i.e., $\rm CrCoNi$, $\rm CrMnCoNi$, $\rm CrFeCoNi$, and $\rm CrMnFeCoNi$), and the three softer ones (i.e., $\rm FeNi$, $\rm Fe_{0.4}Mn_{0.27}Ni_{0.26}Co_{0.05}Cr_{0.02}$, and $\rm Fe_{0.7}Ni_{0.11}Cr_{0.19}$) . Fig.~\ref{fig:roughness}(c) shows the stacking fault area and roughness for all these alloys at their corresponding depinning stress. While all alloys host rough dislocations, the roughness of the four stronger alloys displays overhangs that resemble domain walls in disordered ferromagnetic thin films~\cite{mughal2010effect,bohn2014statistical,laurson2014universality}, that lead to dipolar-forces dominated crossover effects~\cite{laurson2014universality}. It is also worth noting that the observed alloy $\rm CrCoNi$ with the largest depinning stress and visibly large roughness, has the smallest stacking fault area, consistent with experimental evidence on the key role of stacking faults for this alloy~\cite{zhang2020short}. 

\begin{figure*}[tbh]
	\centering
	\includegraphics[width=\linewidth]{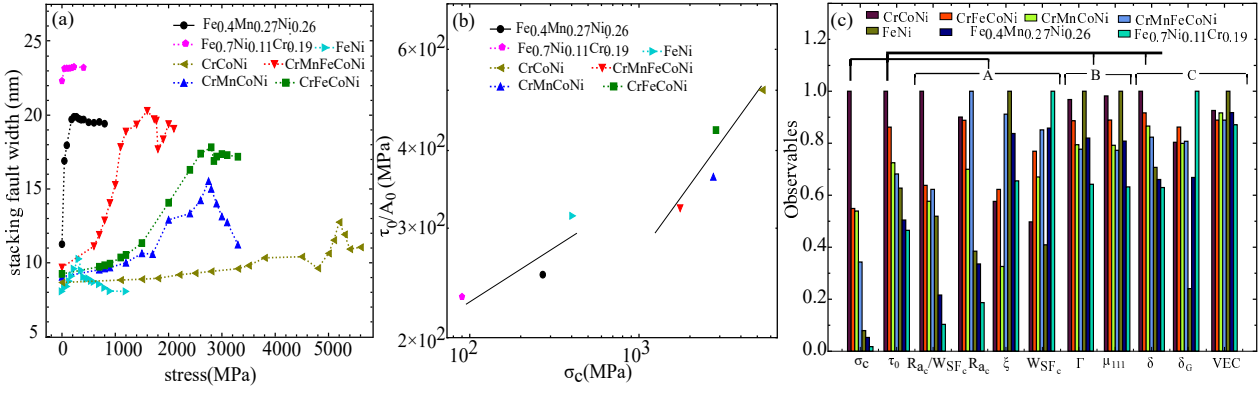}
	\caption{(a) Stacking fault width between two partial dislocation dissociated from an edge dislocation as a function of applied stress for several HEAs. (b)  $\sigma_{c}$ versus $\tau_0/A_0$ (eq. 1). (c) $\sigma_{c}$ (This study), $\tau_0$ (Varvenne's model) and their related descriptors. Descriptors A (i.e. saturation roughness at depinning stress($R_{a_c}$), stacking fault width at depinning stress ($W_{SF_c}$), hardening factor ($R_{a_c}/W_{SF_c}$) and correlation length ($\xi$) ) and Descriptors B (i.e. line tension($\Gamma$) and shear modulus ($\mu_{111/110}$)) were calculated based on MD simulations, while Descriptors C (i.e.  atomic misfit ($\delta$),  shear modulus misfit($\delta_G$) ,  and Valence electron concentration ($VEC$)) were calculated based on Ref.~\cite{kittel1996introduction,martienssen2006springer,guo2011effect}. All data were normalized for each quantity for the different alloys}
	\label{fig:stacking fault}
\end{figure*}

The roughness of a dislocation line is given by\cite{patinet2011,peterffy2020}: $R\left(l\right)={\left<\left(x\left(z+l\right)-x\left(z\right)\right)^{2}\right>}^\frac{1}{2}$
where $x$ and $z$ represent glide and dislocation line directions, respectively (Fig.\ref{fig:roughness}a) and $x(z)$ refers to the dislocation segment position at height z. As ($\sigma\rightarrow\sigma_c$), dislocations relax to new configurations through avalanches~\cite{papanikolaou2017avalanches}. For $\sigma<\sigma_c$, one defines also the Hurst exponent $H$ through~\cite{patinet2011,peterffy2020,geslin2018thermal}:$\log\left(R\right)={H\log\left(l\right)+c}$, with $H$ ranging in [0.5-1] for dislocation lines~\cite{patinet2011,peterffy2020,geslin2018thermal}. Going beyond $\sigma_c$ , the mobility of dislocations are influenced by dislocation-solute interactions. In fcc SSAs, the relationship between the velocity of dislocation line and external stress is~\cite{peterffy2020,zhao2017, papanikolaou2017avalanches}: $v (\sigma) \propto {(\sigma- \sigma_{c})}^\beta$.
 Fig.~\ref{fig:velocity}(a) shows first the position and the velocity of an edge dislocation as a function of time for equimolar FeNi and CrMnFeCoNi in two different $\sigma$s. Fig.~\ref{fig:velocity}(b) represents velocity as a function of  $\sigma - \sigma_{c}$, which is then fitted accordingly. The variation of the exponent values $\beta$ is unexpected and we interpret it as a result of strong collective pinning of the two partials; indeed recently similar physics has been found in magnetic domain walls underlining the importance of collective phenomena \cite{skaugen2021depinning}. It is clear that the effective $\beta$ exponent is higher for the three softer alloys, than for the other four alloys, and meanwhile $\rm FeNi$ and $\rm Fe_{0.7}Ni_{0.11}Cr_{0.19}$ show similar results to previous studies~\cite{peterffy2020,zhao2017}. 

Fig.~\ref{fig:hcp}(a) represents the radial distribution function ($g(r)$) for CrCoNi and CrMnFeCoNi alloys when $\sigma=\sigma_{c}$.  More fluctuations in $g(r)$ can be seen for the strongest alloy(i.e. CrCoNi). While at $\sigma\leq\sigma_c$, the number of hexagonal close packed (hcp) atoms inside the stacking faults approaches a constant value (Fig.~\ref{fig:hcp}(b)), this number increase drastically for $\sigma>\sigma_c$ (Fig.~\ref{fig:hcp}(c) after showing a similar behavior (regime) with Fig.~\ref{fig:hcp}(b). It is noteworthy that the velocity and stacking fault width of dislocation lines (see also Fig.~\ref{fig:stacking fault}(a) below) were reported just for the first regime. At $\sigma>\sigma_{c}$, the roughness was averaged from five configurations of dislocation lines at different times. Due to the high fluctuation of dislocation lines in four stronger alloys and to compare the roughness with the same criteria, the correlation length was calculated by considering H=0.5 (see Fig.~\ref{fig:roughness2}).  
Fig.~\ref{fig:roughness2} shows the correlation length ($\xi$) and saturation roughness ($R_a$) as a function of  stress. Although a large correlation length for the three softer materials was observed, with $R_a$ at depinning for four stronger alloys being much larger than the three softer. 

Fig \ref{fig:stacking fault}(a) shows the stacking fault width ($W_{SF}$) between two partial dislocation lines for each alloy as a function of stress. For $W_{SF}$, we calculate the average location of each partial dislocation line in the glide direction, then subtract the two. We find that the stacking fault width ($W_{SF_c}$), is maximum at the depinning stress point $\sigma_c$.

In commonly adopted models of solid solution strengthening, the dissociation, in face-centered cubic (fcc) materials, of an edge dislocation into two partials, during loading, are minimally considered; While in most cases, the effects are naturally expected to be minimal, they are significant in the case where the roughness of dislocation lines is comparable with the stacking fault width. In particular, it is worth noticing that a solid solution strengthening model for equiatomic alloys~\cite{varvenne2016theory,noehring2019, gludovatz2014fracture,wu2014temperature} represents a relationship between the yield stress at 0K $\tau_{0}$, and only the atom size mismatch ($\delta$)~\cite{varvenne2016theory,noehring2019}, as:
\begin{equation}
    {\tau_{\rm 0}} \sim \left( {\frac{1}{\Gamma}}\right)^\frac{1}{3}\left( {\mu \frac{1+\nu}{1-\nu}}\right)^\frac{4}{3} \left({\delta}\right)^\frac{4}{3}
\label{eq.1}
\end{equation}
where $\Gamma$, $\mu$ , and $\nu$ are line tension, shear modulus, and Poisson ratio, respectively. Our results have been developed on the phenomenological basis of Eq.~\ref{eq.1} and may be considered in relative agreement. Nevertheless, our results make a further step in the investigation of complex, mutual elastic interactions of rough dislocation partials, when their mutual average distance $W_{SF_c}$ is comparable to the partials' roughness $R_{a_c}$. While Eq.~\ref{eq.1} includes minimal effects of mutually parallel partial dislocations  at  stacking faults~\cite{varvenne2016}, we show that there is a non-trivial interplay of very strong pinning disorder (large $R_{a_c}$) and relatively small $W_{SF_c}$.

To  realize which quantities affect depinning stress, correlation between depinning stress and its descriptors (i.e. saturation roughness at depinning stress($R_{a_c}$), stacking fault width at depinning stress ($W_{SF_c}$), hardening factor ($R_{a_c}/W_{SF_c}$) and correlation length ($\xi$) ), Varvenne's model yield stress at 0 K and its descriptors (i.e. line tension($\Gamma$) and shear modulus ($\mu_{111/110}$)), misfit parameters and Valence electron concentration ($VEC$)) is shown in Fig.~\ref{fig:stacking fault}c) for all SSAs when the values are normalized. A good correlation between the depinning stress and model predictions for the yield stress is observed for three alloys with the lowest yield stress (Fig.~\ref{fig:stacking fault}c)), whereas the scaling for the four stronger alloys is different (see Fig.~\ref{fig:stacking fault}(b) with a line to guide the eye).

Fig.~\ref{fig:stacking fault}(c) represents the strong correlation between depinning stress and $\delta$, and the ratio of $R_{a_c}/W_{SF_c}$.  The last correlations represent a novel hardening factor. The stronger alloys at depinning stress have high roughness as well as low stacking fault width, indicating that in the four stronger alloys the difference between misfit parameters (misfit shear modulus and misfit atomic size) in the fcc and hcp phase can play an important role where the lower stacking fault width (less number of hcp atoms) leads to a higher strength.


In summary, this study investigated the geometry of edge dislocations and their mobility under the application of external stress for seven random fcc SSAs. At the top four stronger alloys, due to the fact that stacking fault widths are very small, the mutual elastic interactions of the corresponding partials at $\sigma
\simeq\sigma_c$ are really high and lead to exceptional strength through an interplay of strong-disorder depinning of two closely spaced and spatially correlated elastic lines. Even though elastic depinning theories~\cite{Fisher:1998fk} have not yet investigated this particular regime, we provided extensive and consistent evidence for the existence of this fundamentally novel regime, dominated by roughness-induced strong elastic interactions at the stacking fault that may significantly influence hardening in these materials.

\indent \textit{Acknowledgments}\textemdash
We would like to thank Pawel Sobkowicz for insightful discussions. We acknowledge support from the European Union Horizon 2020 research and innovation program under grant agreement no. 857470 and from the European Regional Development Fund via the Foundation for Polish Science International Research Agenda PLUS program grant No. MAB PLUS/2018/8. We acknowledge the computational resources provided by the High Performance Cluster at the National Centre for Nuclear Research in Poland.

%

\end{document}